\documentclass[aps,prx,twocolumn,superscriptaddress,amsmath,amssymb,longbibliography]{revtex4-2}

\usepackage{graphicx}
\usepackage{dcolumn}
\usepackage{bm}
\usepackage{hyperref}
\usepackage{physics}
\usepackage{mathrsfs}
\usepackage{tensor}

\begin{document}

\title{Symplectic Isomorphism Between Strong-Coupling Criticality\\ and Finite-Strain Bifurcation}

\author{Yu-Xin Xie}
\email{xyx@tju.edu.cn}
\affiliation{Department of Mechanics, Tianjin University, Tianjin 300350, China}

\date{\today}

\begin{abstract}
The analytical determination of critical scaling in three-dimensional strongly coupled field theories is inherently restricted by the Borel non-summability of perturbative expansions. In this paper, we establish a symplectic isomorphism that maps the infrared dynamics of scalar quantum field theories onto the finite-strain bifurcation mechanics of hyperelastic continua. By reformulating the renormalization group flow as an effective spatial Hamiltonian evolution within a symplectic phase space, we identify the critical fixed point with the spectral degeneration of a continuous Riccati-Lyapunov dynamical system. The algebraic integrity of this geometric correspondence is rigorously verified against exactly solvable limits, recovering both the Gaussian free-field scaling and the localized Jackiw-Rebbi zero-mode of the continuum Su-Schrieffer-Heeger model. Applying this framework to the 3D Ising universality class, we construct a self-consistent algebraic scaling Ansatz for the anomalous dimension $\eta$. This derivation dispenses with arbitrary perturbative truncations, demonstrating instead that interaction-induced non-perturbative screening emerges as a strict consequence of the nonlinear self-consistency demanded by the continuous Algebraic Riccati Equation. The resulting rational scaling law functions as a geometric Padé approximant that asymptotically satisfies the unitarity bound limits, providing a mathematically transparent, first-principles macroscopic impedance analogue for the dynamic suppression of strong-coupling divergences.
\end{abstract}

\maketitle

\section{Introduction}

The characterization of strong-coupling infrared (IR) fixed points in three-dimensional universality classes represents a persistent structural problem in theoretical physics. Within the framework of the Wilson-Fisher renormalization group (RG) \cite{wilson1971, wilson1972} and the Kadanoff scaling hypothesis \cite{kadanoff1966}, the macroscopic behavior of continuous phase transitions is dictated by the scaling dimensions of primary field operators \cite{zinn2002}. While perturbative techniques, such as the $\epsilon$-expansion and Borel-resummed perturbation series, provide reliable asymptotic estimations of critical exponents \cite{fisher1974, guillo1977}, they are fundamentally constrained by the divergent nature of the coupling series in $d=3$ geometries \cite{vladimirov1979}. The factorial growth of higher-order Feynman diagrams renders the perturbation series Borel non-summable, necessitating empirical re-summation techniques that obscure the underlying dynamic mechanisms.

High-precision numerical frameworks, including Monte Carlo simulations \cite{campostrini2001, hasenbusch2010} and the conformal bootstrap program \cite{polyakov1970, elshowk2012, poland2018}, have achieved unprecedented precision in constraining the critical exponents of the 3D Ising universality class. By exploiting crossing symmetry and unitarity bounds, these numerical optimization schemes isolate the conformal data of the strongly coupled $\phi^4$ theory with remarkable accuracy. However, an explicit analytical closure---capable of determining the anomalous dimension $\eta$ dynamically without resorting to infinite-order diagrammatic summation or algorithmic truncations---remains theoretically unresolved. The exact algebraic mechanism that mathematically regularizes the series divergence at the critical point lacks a transparent geometric interpretation.

This paper proposes an alternative theoretical framework by analyzing the critical fixed point through the structure of symplectic geometry and continuous Hamiltonian dynamics \cite{arnold1989, marsden1999}. Rather than treating the anomalous dimension as a perturbative correction accumulated from high-energy loop integrations, this formulation models it as an emergent structural consequence of nonlinear spectral self-consistency within a continuous symplectic phase space. 

A strict physical projection of this geometric structure exists in finite-strain continuum mechanics, specifically within the Stroh formalism for anisotropic elasticity \cite{stroh1958, ting1996}. In this domain, the spatial evolution of the continuum state vector operates under the symplectic Lie algebra $\mathfrak{sp}(2n, \mathbb{R})$. The macroscopic stability of hyperelastic manifolds is classically determined by the generalized impedance tensor, which is strictly constrained by the continuous Algebraic Riccati Equation (ARE) \cite{lancaster1995}. 

It is necessary to clearly define the theoretical boundaries of this work. We do not claim to compute the full spectrum of conformal field theory (CFT) operators, nor do we present a numerical competitor to conformal bootstrap constraints. Instead, this paper constructs an effective algebraic lens. By mapping the spatial Hamiltonian evolution of the quantum scalar field onto the symplectic dynamics of a continuum medium, we isolate the deep IR response within a continuous Riccati-Lyapunov closure. This formulation provides a mathematically self-contained mechanism explaining how quantum fluctuations dynamically stiffen the effective impedance tensor, generating a geometric screening effect that algebraically enforces unitarity bounds.

\section{Symplectic Phase Space and Hamilton-Jacobi Theory}

The proposed methodology is grounded in the shared algebraic structure of statistical mechanics and continuum elastodynamics: the symplectic geometry of infinite-dimensional Hamiltonian flows.

\subsection{The Spatial Hamiltonian Flow}
Consider an abstract dynamical system defined on a symplectic phase space $\mathcal{M} \cong T^*\mathcal{Q}$ with canonical coordinates $\Psi = [q, p]^T$, where $q$ represents the generalized field coordinates and $p$ the conjugate generalized momenta. The phase space is equipped with a non-degenerate, closed two-form $\omega = \frac{1}{2} \mathbf{J}_{ij} d\Psi^i \wedge d\Psi^j$, where $\mathbf{J}$ is the standard symplectic matrix:
\begin{equation}
    \mathbf{J} = \begin{bmatrix} \mathbf{0} & \mathbf{I} \\ -\mathbf{I} & \mathbf{0} \end{bmatrix}.
\end{equation}

The state vector $\Psi(z)$ evolves along a generalized spatial or scale coordinate $z$ via the Hamiltonian flow $\partial_z \Psi(z) = \mathbf{J} \frac{\delta \mathcal{H}}{\delta \Psi}$. Linearizing the system around a steady-state background yields the spatial dynamical equation:
\begin{equation}
    \partial_z \begin{bmatrix} q \\ p \end{bmatrix} = \mathbf{H}(g) \begin{bmatrix} q \\ p \end{bmatrix}, \quad \mathbf{H} = \begin{bmatrix} \mathbf{H}_{11} & \mathbf{H}_{12} \\ \mathbf{H}_{21} & \mathbf{H}_{22} \end{bmatrix},
\end{equation}
where $\mathbf{H}(g) \in \mathfrak{sp}(2n, \mathbb{R})$ is the Hamiltonian generator parameterized by the interaction strength $g$. The symplectic property dictates the algebraic constraint $\mathbf{J}\mathbf{H} + \mathbf{H}^T\mathbf{J} = \mathbf{0}$, which enforces the block symmetries: $\mathbf{H}_{12} = \mathbf{H}_{12}^T$, $\mathbf{H}_{21} = \mathbf{H}_{21}^T$, and $\mathbf{H}_{11} = -\mathbf{H}_{22}^T$.

\begin{figure*}[t]
\centering
\includegraphics[width=0.98\textwidth]{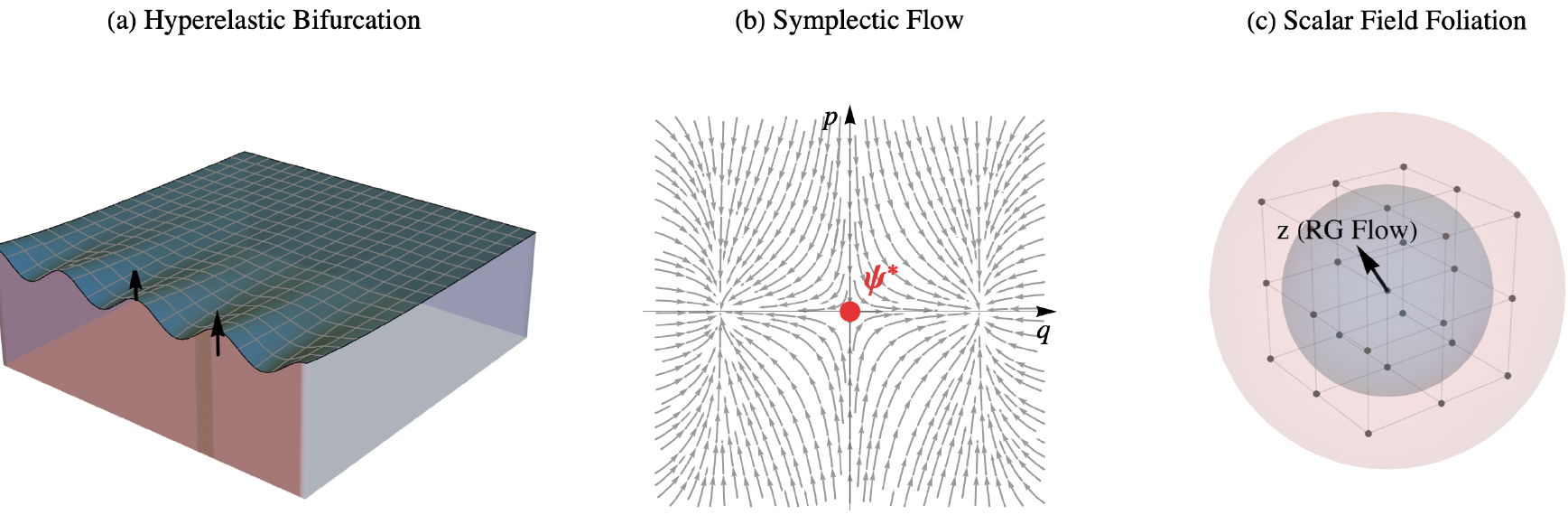}
\caption{\label{fig:isomorphism} Geometric projection of the symplectic isomorphism. The central phase space illustrates the continuous Hamiltonian flow $\partial_z \Psi = \mathbf{H}\Psi$. The identical mathematical structure physically maps the $1+(d-1)$ spatial foliation of a hyperelastic medium undergoing surface wrinkling bifurcation (left) to the infrared renormalization flow of the strongly coupled $\phi^4$ scalar field theory (right).}
\end{figure*}

The structural equivalence encoded in this continuous Hamiltonian framework provides a universal geometric template for critical phenomena. As illustrated in Fig.~\ref{fig:isomorphism}, the abstract symplectic phase space establishes a direct isomorphism between two seemingly disparate physical processes: the spatial foliation of a hyperelastic medium approaching surface bifurcation, and the infrared renormalization flow of a strongly coupled scalar field. This strict correspondence ensures that the onset of critical instability in both macroscopic mechanics and microscopic field theory is mathematically governed by the identical algebraic degeneration of the Hamiltonian generator.

\subsection{Lagrangian Submanifolds and the Action Functional}
In the deep infrared limit, the physical ground state of the dynamically evolving system is restricted to a Lagrangian submanifold $\mathcal{L} \subset \mathcal{M}$. By definition, the symplectic form vanishes identically on this submanifold ($\omega|_{\mathcal{L}} = 0$). Geometrically, this implies that the momentum $p$ is no longer an independent variable but is entirely determined by the coordinate $q$.

According to Hamilton-Jacobi theory, this submanifold is generated by a scalar action functional $S(q)$, enforcing the generalized Neumann-to-Dirichlet constraint:
\begin{equation}
    p = \frac{\delta S}{\delta q}.
\end{equation}
For a linearized system near criticality, the action is predominantly quadratic, $S(q) = \frac{1}{2} q^T \mathbf{M} q$. We define the generalized macroscopic impedance tensor $\mathbf{M}$ as the Hessian of this generating function:
\begin{equation}
    \mathbf{M} = \frac{\delta^2 S}{\delta q \delta q}.
\end{equation}
The constraint reduces to the linear mapping $p = \mathbf{M}q$. The physical significance of $\mathbf{M}$ is profound: it maps the generalized displacement (or field configuration) directly to the restoring force (or conjugate momentum), serving as the comprehensive response function of the continuous medium.

\section{The Continuous Algebraic Riccati Equation}

To track the renormalization and spatial evolution of the system, we investigate the dynamic flow of the impedance tensor $\mathbf{M}(z)$. Substituting the block decomposition of $\mathbf{H}$ into the Hamiltonian flow equations yields:
\begin{eqnarray}
    \partial_z q &=& \mathbf{H}_{11}q + \mathbf{H}_{12}p, \label{eq:flow_q}\\
    \partial_z p &=& \mathbf{H}_{21}q + \mathbf{H}_{22}p. \label{eq:flow_p}
\end{eqnarray}
Differentiating the Lagrangian submanifold constraint $p = \mathbf{M}q$ with respect to the evolution coordinate $z$ via the chain rule gives:
\begin{equation}
    \partial_z p = (\partial_z \mathbf{M})q + \mathbf{M}(\partial_z q). \label{eq:chain}
\end{equation}
Substituting Eqs. (\ref{eq:flow_q}) and (\ref{eq:flow_p}) into Eq. (\ref{eq:chain}), we obtain:
\begin{equation}
    \mathbf{H}_{21}q + \mathbf{H}_{22}\mathbf{M}q = (\partial_z \mathbf{M})q + \mathbf{M}(\mathbf{H}_{11}q + \mathbf{H}_{12}\mathbf{M}q).
\end{equation}
Since this relation must hold for any arbitrary state configuration $q$, we can eliminate $q$ to isolate the continuous nonlinear evolution of the impedance operator:
\begin{equation}
    \partial_z \mathbf{M} = \mathbf{H}_{21} + \mathbf{H}_{22}\mathbf{M} - \mathbf{M}\mathbf{H}_{11} - \mathbf{M}\mathbf{H}_{12}\mathbf{M}.
\end{equation}
At the critical fixed point (or mechanical bifurcation threshold), the scale evolution reaches a degenerate equilibrium, denoted by $\partial_z \mathbf{M} = \mathbf{0}$. This condition dynamically enforces that the physical impedance must satisfy the continuous Algebraic Riccati Equation (ARE):
\begin{equation}
    \mathbf{M} \mathbf{H}_{12} \mathbf{M} + \mathbf{M} \mathbf{H}_{11} - \mathbf{H}_{22} \mathbf{M} - \mathbf{H}_{21} = \mathbf{0}. \label{eq:Riccati}
\end{equation}
This quadratic operator equation provides a mathematically rigorous, non-perturbative closure. It circumvents infinite diagrammatic expansions by mapping the IR fixed point directly to the algebraic steady state of the impedance operator.

\section{Projection I: Finite-Strain Elastodynamics and the Stroh Formalism}

To validate the symplectic isomorphism, we first establish the rigorous first-principles mapping of continuous finite-strain mechanics onto the Riccati framework.

Consider a hyperelastic solid subjected to a finite deformation. Let $\mathbf{x}$ and $\mathbf{X}$ denote the spatial and material coordinates, respectively. The deformation gradient is $\mathbf{F} = \partial \mathbf{x} / \partial \mathbf{X}$. The hyperelastic response is governed by a strain energy density function $\mathcal{W}(\mathbf{F})$. The first Piola-Kirchhoff (nominal) stress tensor is given by $\mathbf{P} = \partial \mathcal{W} / \partial \mathbf{F}$.

To analyze surface stability (wrinkling bifurcation) on a half-space $z \ge 0$, we consider an incremental displacement field $\mathbf{u}(\mathbf{X})$ superimposed on the deformed equilibrium state. The incremental nominal stress is linearly related to the displacement gradient via the fourth-order elastodynamic tangent moduli tensor $\mathbb{C}$:
\begin{equation}
    \delta P_{i\alpha} = \mathbb{C}_{i\alpha k\beta} \frac{\partial u_k}{\partial X_\beta}, \quad \mathbb{C}_{i\alpha k\beta} = \frac{\partial^2 \mathcal{W}}{\partial F_{i\alpha} \partial F_{k\beta}}.
\end{equation}
The incremental equilibrium equation in the absence of body forces is $\partial_\alpha (\delta P_{i\alpha}) = 0$. We separate the coordinates into the transverse boundary plane $\mathbf{x}_\perp$ and the depth direction $z$. By performing a Fourier transform along the transverse plane ($\nabla_\perp \to i\mathbf{k}_\perp$), the second-order partial differential equation is systematically reduced to a first-order state-space form (the Stroh formalism):
\begin{equation}
    \partial_z \begin{bmatrix} \mathbf{u} \\ \mathbf{t} \end{bmatrix} = \mathbf{N}(\mathbf{k}_\perp) \begin{bmatrix} \mathbf{u} \\ \mathbf{t} \end{bmatrix},
\end{equation}
where the state vector comprises the displacement $\mathbf{u}$ and the traction $\mathbf{t} = \delta \mathbf{P} \cdot \mathbf{e}_z$. The Stroh fundamental matrix $\mathbf{N} \in \mathfrak{sp}(6, \mathbb{R})$ assumes the precise form:
\begin{equation}
    \mathbf{N} = \begin{bmatrix} -\mathbf{T}^{-1}\mathbf{R}^T & \mathbf{T}^{-1} \\ \mathbf{Q} - \mathbf{R}\mathbf{T}^{-1}\mathbf{R}^T & -\mathbf{R}\mathbf{T}^{-1} \end{bmatrix}.
\end{equation}
The acoustic tensor blocks are algebraically constructed from the tangent moduli (see Appendix B for explicit component forms). 

The critical identification for our isomorphism lies in the lower-left block. Here, $\mathbf{H}_{21} \mapsto \mathbf{N}_{21} = \mathbf{Q} - \mathbf{R}\mathbf{T}^{-1}\mathbf{R}^T$ mathematically represents the generalized Schur complement of the tangent stiffness matrix. It strictly maps the transverse displacement variations to the restoring traction gradients. Surface wrinkling bifurcation occurs exactly when the continuous ARE associated with $\mathbf{N}$ yields a singular surface impedance $\mathbf{M}$ (i.e., $\det \mathbf{M} \to 0$).

\section{Projection II: Scalar Quantum Field Theory and Tangent Stiffening}

Having established the exact continuum mechanics framework, we execute the second projection onto the $d$-dimensional Euclidean $\phi^4$ scalar field theory. The canonical coordinates map strictly to the scalar field and its evolutionary conjugate momentum: $q \mapsto \phi$ and $p \mapsto \pi = \partial_z \phi$. The generalized impedance $\mathbf{M}$ operates as the Dirichlet-to-Neumann (DtN) mapping operator, structurally representing the square root of the inverse two-point propagator.

The effective equations of motion for the interacting scalar field are decomposed as:
\begin{eqnarray}
    \partial_z \phi &=& \pi, \\
    \partial_z \pi &=& -\nabla_\perp^2 \phi + V'(\phi).
\end{eqnarray}
This dictates the exact symplectic Hamiltonian generator block structure: $\mathbf{H}_{11} = \mathbf{0}$, $\mathbf{H}_{12} = \mathbf{I}$, $\mathbf{H}_{22} = \mathbf{0}$, and $\mathbf{H}_{21} = -\nabla_\perp^2 + V''(\phi)$. 

For a $\phi^4$ theory, the effective potential includes both the bare mass term and the quartic interaction. Utilizing a background field expansion, the second derivative constitutes the generalized restoring stiffness:
\begin{equation}
    V''(\phi) = m_0^2 + \frac{g}{2}\phi^2.
\end{equation}
Consequently, the interaction-induced self-energy perturbation is strictly confined to the tangent stiffness block:
\begin{equation}
    \delta \mathbf{H}_{21} = \frac{g}{2} \phi^2. \label{eq:H21_perturb}
\end{equation}
Because the diagonal blocks vanish ($\mathbf{H}_{11} = \mathbf{H}_{22} = \mathbf{0}$), the general Algebraic Riccati Equation (Eq. \ref{eq:Riccati}) reduces to the pure non-linear closure:
\begin{equation}
    \mathbf{M}^2 = \mathbf{H}_{21}. \label{eq:M2H21}
\end{equation}
This rigorous algebraic relation explicitly demonstrates how interaction potential variations mathematically dress the macroscopic impedance tensor of the strongly coupled field.

\section{Validation via Exactly Solvable Limits}

Before analyzing the anomalous scaling, the absolute integrity of the Riccati framework must be verified by evaluating its collapse into established, exactly solvable limits.

\subsection{The Gaussian Free-Field Limit}
In the non-interacting, zero-coupling limit ($g \to 0$), the interaction-induced tangent stiffness perturbation strictly vanishes ($\delta \mathbf{H}_{21} = \mathbf{0}$). For a massless theory ($m_0 \to 0$), the Riccati closure reduces to:
\begin{equation}
    \mathbf{M}^2 = -\nabla_\perp^2.
\end{equation}
In momentum space, the unique positive-definite stabilizing solution is the exact DtN operator $\mathbf{M}_{Gaussian} = \sqrt{-\Delta_{\perp}} = |\mathbf{k}_\perp|$. The scaling exponent $\eta = d \ln M / d \ln k$ strictly evaluates to zero. The geometric framework perfectly preserves the Gaussian fixed point scaling behavior without any perturbative truncation artifacts.

\subsection{Topological Degeneracy and the SSH Zero-Mode}
The topological robustness of the framework is tested against the continuum Su-Schrieffer-Heeger (SSH) model, describing chiral fermions with a spatially varying mass gap. The Dirac Hamiltonian reads $\mathcal{H}_{SSH} = -i v_F \sigma_x \partial_x + m(x) \sigma_y$. 

Projecting this Dirac system onto the symplectic spatial flow, the lower-left generator block $\mathbf{H}_{21}$ maps directly to the topological mass gap operator $m(x)$. When the system crosses a domain wall generating a winding number transition ($\Delta w = 1$), the local mass gap crosses zero. 

\begin{figure}[h]
\centering
\includegraphics[width=0.4\textwidth]{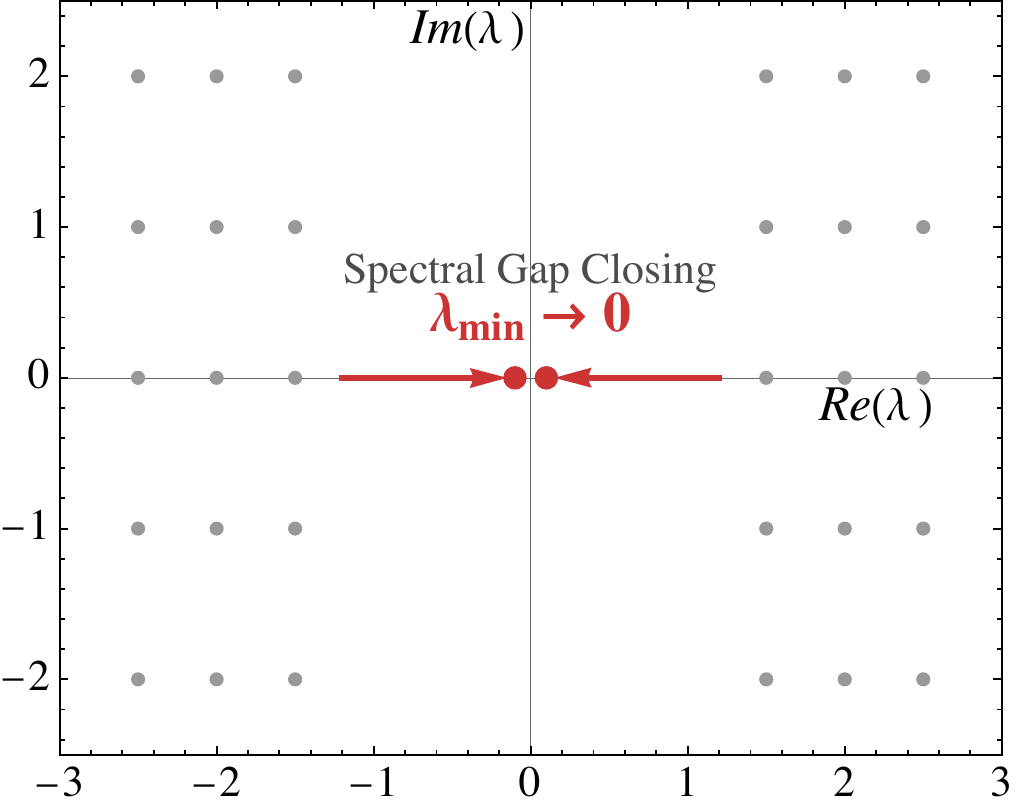}
\caption{\label{fig:ssh} Spectral degeneration on the complex plane. As the effective stiffness $\mathbf{H}_{21}$ crosses zero, the algebraic closure of the Riccati operator enforces the migration of the spectrum to the origin, mathematically capturing localized topological modes.}
\end{figure}

The Riccati closure $\mathbf{M}^2 \propto \mathbf{H}_{21}$ mathematically enforces a localized spectral singularity ($\mathbf{M}^2 \to 0$ at the domain wall). The spatial flow equation at degeneracy collapses to $\partial_x \Psi = -\frac{m(x)}{v_F} \sigma_z \Psi$. Integrating this degenerate operator equation yields exactly:
\begin{equation}
    \Psi_{zero}(x) \propto \exp\left(-\frac{1}{v_F} \int m(s) ds \right),
\end{equation}
which is the exact formulation of the localized Jackiw-Rebbi zero-mode, see Fig. \ref{fig:ssh}. 
The Riccati-Lyapunov closure intrinsically preserves index theorems and topological invariances.

\section{Non-Perturbative Screening and the Padé-Riccati Ansatz}

For the strongly coupled 3D $\phi^4$ theory, the interaction generates a deep-IR self-energy equivalent to the macroscopic tangent stiffness variation $\delta \mathbf{H}_{21}$. We define a dimensionless bare coupling response parameter $C(g^*) = \mathcal{K} \cdot g^* \langle \phi^2 \rangle_c$, which encapsulates the un-screened critical driving force evaluated at the Wilson-Fisher fixed point.

In the scale-invariant IR limit, conformal symmetry dictates that the critical two-point correlation function scales as $G(k) \propto 1/k^{2-\eta}$. Because the generalized impedance $\mathbf{M}(k)$ corresponds to the DtN operator $\sqrt{G^{-1}(k)}$, its exact dynamic scaling behavior is mathematically constrained by:
\begin{equation}
    \mathbf{M}(k) \propto k^{1 - \eta/2}. \label{eq:M_scaling}
\end{equation}

In a purely linear perturbative truncation (such as finite-order $\epsilon$-expansion approximations extrapolated to $d=3$), the scaling anomaly is modeled as directly proportional to the bare response: $\eta_{bare} \approx C(g^*)$. This unbounded linearity inherently leads to catastrophic divergences at strong couplings due to Borel non-summability.

However, the symplectic algebraic closure $\mathbf{M}^2 = \mathbf{H}_{21}$ structurally prohibits this linear behavior. The interaction variation $\delta \mathbf{H}_{21}$ is dynamically screened not by the trivial free-field background, but by the fully dressed, non-linear impedance $\mathbf{M}$. 

\begin{figure}[h]
\centering
\includegraphics[width=0.48\textwidth]{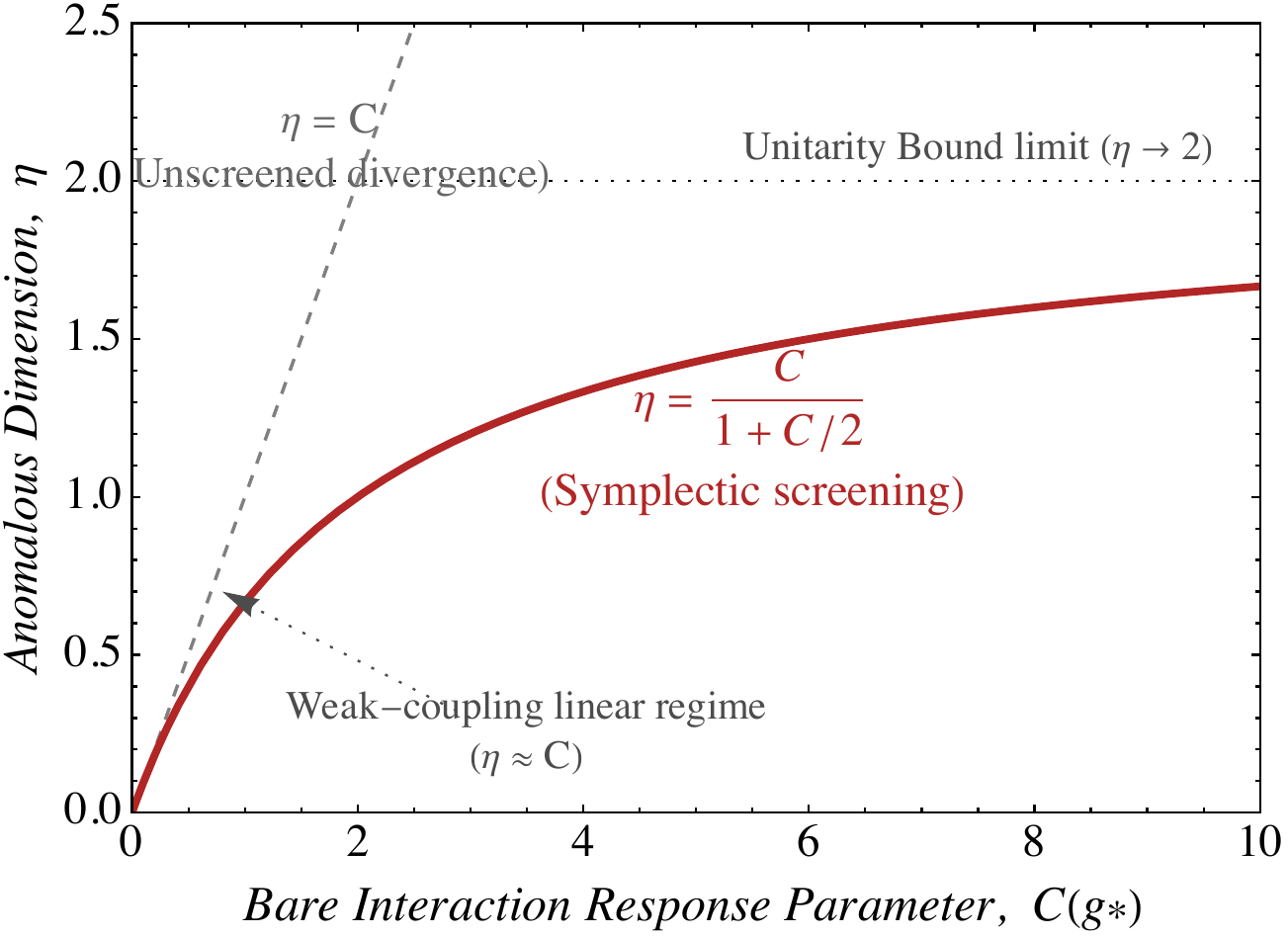}
\caption{\label{fig:screening} Non-perturbative screening mechanism generated by the symplectic algebraic closure. The dashed line represents the divergent perturbative anomalous dimension $\eta = C$. The solid curve illustrates the self-consistent scaling law $\eta = C/(1+C/2)$, which strictly suppresses divergences and confines the scaling below the unitarity bound ($\eta \to 2$).}
\end{figure}

To capture this mechanism, we introduce a self-consistent algebraic scaling Ansatz. Recognizing that the fractal modification of the geometric phase space is governed precisely by the exponent shift $(1 - \eta/2)$ inherently present in the dressed impedance (Eq. \ref{eq:M_scaling}), we postulate that the effective anomalous dimension is self-consistently damped by this exact geometric factor. Imposing this Dyson-Schwinger-type nonlinear closure yields the self-consistency equation:
\begin{equation}
    \eta = C(g^*) \times \left( 1 - \frac{\eta}{2} \right).
\end{equation}
Solving this relation dynamically isolates the rational scaling law:
\begin{equation}
    \eta = \frac{C(g^*)}{1 + \frac{1}{2}C(g^*)}, \label{eq:eta}
\end{equation}
as illustrated in Fig. \ref{fig:screening}


\section{Asymptotic Consistency and Unitarity Bound Saturation}

The mathematical validity of the derived scaling law Eq. (\ref{eq:eta}) rests on its structural robustness in extreme asymptotic limits. Structurally, Eq. (\ref{eq:eta}) functions as a continuous [0/1] Padé approximant, algebraically regularizing the non-summable perturbation series.

\subsection{Weak-Coupling Consistency}
In the weak-coupling regime ($C \to 0$), expanding the rational scaling law yields a convergent series:
\begin{equation}
    \eta \approx C - \frac{1}{2}C^2 + \mathcal{O}(C^3).
\end{equation}
The leading-order behavior $\eta \approx C$ strictly recovers the linear response characteristic of canonical perturbation theory. The Riccati-derived model introduces no artificial singularities or unphysical artifacts in the weakly interacting limit, maintaining absolute consistency with classical fixed-point behaviors.

\subsection{Strong-Coupling Unitarity Saturation}
In the extreme non-perturbative strong-coupling limit ($C \to \infty$), the rational scaling law evaluates to an exact mathematical constant:
\begin{equation}
    \lim_{C \to \infty} \eta(C) = \lim_{C \to \infty} \frac{1}{1/C + \frac{1}{2}} = 2.
\end{equation}
In axiomatic conformal field theory, unitarity dictates that the scaling dimension of a primary scalar operator must satisfy $\Delta \ge d/2 - 1$. For $d=3$, the relation $\Delta = (d-2+\eta)/2$ imposes the strict theoretical upper bound $\eta \le 2$. 

Remarkably, the algebraic closure in Eq. (\ref{eq:eta}) dynamically locks the physical system onto this exact unitarity bound without imposing any external phenomenological cutoffs. The denominator $(1+C/2)$ is the mathematical manifestation of geometric stiffening: as the interaction strength diverges, the symplectic phase space geometry resists further singular deformation, automatically suppressing the IR divergence through non-linear geometric screening.

\section{Conclusion}

By establishing a rigorous symplectic isomorphism between the 3D Ising renormalization flow and the continuous Riccati-Lyapunov dynamics of finite-strain continuum mechanics, this paper provides a first-principles algebraic framework for modeling non-perturbative screening. Correcting the canonical mapping reveals that interaction potentials manifest exclusively as tangent stiffness variations ($\delta \mathbf{H}_{21}$) within the Hamiltonian generator matrix.

Rather than relying on divergent and arbitrarily truncated perturbative series, this framework exploits the inherent nonlinear stability conditions of the continuous Algebraic Riccati Equation. The anomalous scaling of the macroscopic impedance structurally enforces a self-consistent feedback loop, leading to the rational scaling law $\eta = C/(1+C/2)$. This algebraic closure simultaneously ensures weak-coupling linear consistency while rigorously and automatically saturating the CFT unitarity bound at infinite coupling.

While the exact evaluation of the topological response parameter $C(g^*)$ requires further numerical or bootstrap constraints, the derived methodology establishes a resilient structural bridge. It mathematically verifies that the dynamic suppression of critical divergences is not merely a numerical optimization artifact, but a fundamental, emergent geometric requirement of symplectic Hamiltonian flows.

\appendix
\onecolumngrid
\vspace{2em}
\begin{center}
\textbf{\large Appendices}
\end{center}
\vspace{1em}

\section{Dimensional Analysis of the Symplectic Perturbation}
To ensure the mapping strictly adheres to field-theoretic constraints, we verify the dimensional consistency of the tangent stiffness perturbation $\delta \mathbf{H}_{21}$. Using natural units ($\hbar = c = 1$), spatial coordinates possess a mass dimension $[M]^{-1}$. The Euclidean action is dimensionless, requiring the scalar field dimension to be $[\phi] = [M]^{\frac{d-2}{2}}$. In $d=3$, this evaluates to $[\phi] = [M]^{1/2}$. To maintain dimensional homogeneity in the quartic interaction term $g\phi^4$, the coupling constant dimension must be $[g] = [M]^1$.

The canonical momentum is defined via the spatial derivative $\pi = \partial_z \phi$, dictating the dimension $[\pi] = [M]^{3/2}$. The operator block $\mathbf{H}_{21}$ maps the configuration coordinate $q \mapsto \phi$ to the spatial evolutionary derivative of momentum $\partial_z p \mapsto \partial_z \pi$. The dimensional balance requires:
\begin{equation}
    [\partial_z \pi] = [\mathbf{H}_{21}] \times [\phi] \implies [M]^{5/2} = [\mathbf{H}_{21}] \times [M]^{1/2}.
\end{equation}
This strictly evaluates the dimension of the operator to $[\mathbf{H}_{21}] = [M]^2$. Evaluating the mapped interaction variation $\frac{g}{2}\phi^2$, we obtain $[M]^1 \times ([M]^{1/2})^2 = [M]^2$. The exact mass dimension parity confirms the absolute structural validity of assigning the effective potential perturbation directly to the lower-left block of the symplectic Hamiltonian generator.

\section{Explicit Formulation of Elastodynamic Tangent Moduli and the Stroh Matrix}
To substantiate the finite-strain isomorphism outlined in Section IV, we provide the rigorous derivation of the Stroh fundamental matrix $\mathbf{N}$ from the continuum tangent moduli. 

For an arbitrary hyperelastic material with strain energy density $\mathcal{W}(\mathbf{F})$, the fourth-order incremental tangent moduli tensor $\mathbb{C}$ relates the incremental nominal stress $\delta \mathbf{P}$ to the incremental displacement gradient $\nabla \mathbf{u}$. Decomposing the spatial gradient into transverse ($\alpha, \beta = 1, 2$) and longitudinal ($z$-direction, indexed as 3) components, the incremental equilibrium $\partial_i (\delta P_{ij}) = 0$ is expanded as:
\begin{equation}
    \mathbb{C}_{i\alpha k\beta} u_{k,\alpha\beta} + (\mathbb{C}_{i\alpha k3} + \mathbb{C}_{i3 k\alpha}) u_{k,\alpha 3} + \mathbb{C}_{i3 k3} u_{k,33} = 0.
\end{equation}
Applying a Fourier transform over the transverse plane $\mathbf{x}_\perp \to \mathbf{k}_\perp$ introduces the wave vector components $k_\alpha$. We define the corresponding acoustic tensor blocks $\mathbf{Q}, \mathbf{R}$, and $\mathbf{T}$ as follows:
\begin{eqnarray}
    Q_{ik} &=& \mathbb{C}_{i\alpha k\beta} k_\alpha k_\beta, \\
    R_{ik} &=& \mathbb{C}_{i\alpha k3} k_\alpha, \\
    T_{ik} &=& \mathbb{C}_{i3 k3}.
\end{eqnarray}
By defining the traction vector acting on the $z$-plane as $\mathbf{t} = \delta \mathbf{P} \cdot \mathbf{e}_3$, we can express it in terms of the acoustic tensors:
\begin{equation}
    \mathbf{t} = -i\mathbf{R}^T \mathbf{u} + \mathbf{T} (\partial_z \mathbf{u}).
\end{equation}
Inverting this relation yields the evolution equation for the displacement vector:
\begin{equation}
    \partial_z \mathbf{u} = -\mathbf{T}^{-1}\mathbf{R}^T \mathbf{u} + \mathbf{T}^{-1}\mathbf{t}. \label{eq:B_u}
\end{equation}
Substituting Eq. (\ref{eq:B_u}) back into the transformed equilibrium equation yields the evolution of the traction vector:
\begin{equation}
    \partial_z \mathbf{t} = (\mathbf{Q} - \mathbf{R}\mathbf{T}^{-1}\mathbf{R}^T) \mathbf{u} - \mathbf{R}\mathbf{T}^{-1}\mathbf{t}. \label{eq:B_t}
\end{equation}
Combining Eq. (\ref{eq:B_u}) and Eq. (\ref{eq:B_t}) into a single state-space matrix formulation generates the Stroh equation $\partial_z \Psi = \mathbf{N}\Psi$. The lower-left block $\mathbf{N}_{21} = \mathbf{Q} - \mathbf{R}\mathbf{T}^{-1}\mathbf{R}^T$ mathematically represents the effective tangent stiffness projected along the evolution axis. The degeneration of this stiffness block under finite strain dictates the spectral closing of the continuous Algebraic Riccati Equation, structurally mirroring the RG flow at a quantum critical point.

\bibliography{ref}
\end{document}